# Anisotropic Landau level splitting and Lifshitz transition induced magnetoresistance enhancement in $ZrTe_5$ crystals

L. Zhou[1], A. Ramiere[1], P. B. Chen[1], J. Y. Tang[1], Y. H. Wu[1], X. Lei[1], G. P. Guo[2], J. Q. He[1,*], and H. T. He[1,†]

[1] *Department of Physics, Southern University of Science and Technology, Shenzhen 518055, China*

[2]*Key Laboratory of Quantum Information, CAS University of Science and Technology of China, Hefei 230026, China*

**Abstract.** Magneto-transport study has been performed in $ZrTe_5$ single crystals. The observed Shubnikov–de Hass quantum oscillation at low temperature clearly demonstrates the existence of a nontrivial band with small effective mass in $ZrTe_5$. Furthermore, we also revealed the 3D anisotropic nature of high-field Landau level splitting in $ZrTe_5$, very different from the 2D behavior measured in previous transport studies. Besides these, an abnormal large enhancement of magnetoresistance appears at high temperatures, which is believed to arise from the Lifshitz transition induced two-carrier transport in $ZrTe_5$. Our study provides more understanding of the physical properties of $ZrTe_5$ and sheds light on potential application of $ZrTe_5$ in spintronics.

[*] hejq@sustc.edu.cn
[†] heht@sustc.edu.cn

Dirac semimetals (DSMs) are three dimensional analogs of graphene, featuring gapless Dirac cones in the bulk states [1]. The well-studied DSMs, such as $Na_3Bi$ and $Cd_3As_2$, have been shown to exhibit ultra-high mobility, extremely large positive magnetoresistance (MR), surface Fermi arc states, and a novel negative MR related with the chiral anomaly in high energy physics [2-7]. Therefore, DSMs have triggered extensive studies exploring the potential applications in spintronics, such as magnetic sensors, as well as the intriguing Dirac fermion physics [1].

Among various DSM candidates, $ZrTe_5$ is of particular interest. It has a layered crystal structure, which means that thickness-dependent physics can be easily accessed, even down to the ultrathin limit, *i.e.*, one monolayer of $ZrTe_5$ [8, 9]. Besides this, $ZrTe_5$ has been shown to exhibit many intriguing phenomena, such as the chiral anomaly induced negative magnetoresistance (MR) [10], the nontrivial Berry phase revealed in quantum oscillation experiment [11], the anomalous Hall effect even with the magnetic field in-plane [12], and possible emergence of density wave states in the quantum limit [13, 14]. What makes $ZrTe_5$ more special is that it lies at the boundary between weak and strong topological insulators [15]. Whether $ZrTe_5$ is a real DSM or not is still under active investigation in the literature [16,17,18].

In this work, we have performed systematic magneto-transport study of high-quality $ZrTe_5$ single crystals. Clear quantum MR oscillations were observed at low temperatures, revealing the existence of a nontrivial band with tiny effective mass in $ZrTe_5$. In the high field region, we also observed well-resolved Landau level splitting. It shows a very strong dependence on the tilting angle of the magnetic field, indicating the 3D anisotropic nature of it, very different from the 2D nature claimed in previous transport studies [13]. Besides these, we also revealed an abnormal large enhancement of MR up to 820% at high temperatures, which was interpreted in terms of the Lifshitz transition induced two-carrier transport in $ZrTe_5$ and might be important for potential application of $ZrTe_5$ in spintronics.

The $ZrTe_5$ single crystals we studied in this work were grown by the CVT method. Stoichiometric proportions of high purity Zr (GRINM 99.995%) and Te (Alfa Aesar 99.999%) were loaded in an evacuated quartz tube at a pressure below $3\times10^{-5}$ Pa and hold at 480 °C for 12 days. The solid-state reaction created polycrystalline $ZrTe_5$ that was hand-milled, then loaded in a new evacuated quartz tube with 5 mg/cm$^3$ of iodine, and finally placed in a horizonal dual zone furnace with a temperature gradient 430-520 °C for two weeks to grow the single crystals. $ZrTe_5$ has a layered orthorhombic crystal structure, with the 2D layers stacked in the *b*-axis. Each 2D layer consists of prismatic $ZrTe_6$ chains along the $a$-axis linked by zigzag chains of Te atoms in the $c$-axis. Therefore, the obtained $ZrTe_5$ crystals are usually needle-like and preferably grown along the $a$-axis. To investigate the magneto-transport properties of these $ZrTe_5$ crystals, six-terminal Hall devices were fabricated with gold wires attached by silver paint, as schematically shown in the inset of Fig. 1. The devices were then measured in an Oxford TeslatronPT system with the base temperature and highest magnetic field of 1.6 K and 14 T, respectively. A lock-in technique was implemented to improve the measurement accuracy. The ac amplitude and frequency of the measuring current is 1 mA and 17 Hz, respectively.

Fig. 1 shows the temperature ($T$) dependence of resistance ($R$) of the bulk $ZrTe_5$ sample. The resistance firstly increases with $T$ decreasing from room temperature. But below 138 K, it shows a metallic behavior. A resistance peak thus appears at $T_p$=138 K, similar to previous studies of $ZrTe_5$ [11, 13]. Note that the values of $T_p$ can vary in a wide temperature range, depending on the thickness or carrier doping of $ZrTe_5$ [8, 19]. The emergence of such a peak anomaly is recently ascribed to the Lifshitz transition in $ZrTe_5$ [17-19].

We then investigated the magnetoresistance (MR) with the magnetic field ($B$) in the ***b*** direction, as shown in Fig. 2 (a). At $T$= 1.6 K, clear Shubnikov–de Hass (SdH) oscillations are observed in low fields, indicating the formation of quantized Landau levels (LLs) in our sample. Also noticed is the huge MR peak appearing around 8 T, accompanied by a dramatic drop of MR around 12 T. Although previous study has explained this phenomenon in terms of the spin density wave in the quantum limit [13], direct evidence is still lacking. To examine the quantum oscillations more clearly, we subtract the smooth MR background from each $R(B)$ curve in Fig. 2 (a). The obtained $\Delta R$ has been plotted as a function of $1/B$ at each temperature, as shown in Fig. 2 (b). For clarity, the $\Delta R(1/B)$ curves have been offset vertically. As expected for the SdH oscillation [11, 13], the MR oscillations in Fig. 2 (b) are periodic in $1/B$ and gradually fade out with increasing temperatures.

The Landau quantization condition is given by the Onsager relation $A_F \frac{\hbar}{eB} = 2\pi\left(n + \frac{1}{2} - \frac{\phi_B}{2\pi} - \delta\right) = 2\pi(n+\gamma)$, where $A_F$ is the extremal cross-sectional area of Fermi surface perpendicular to the field, $\hbar$ is the Plank's constant, $e$ is the electronic charge, $n$ is the Landau level index, $\phi_B$ is a geometrical phase, also known as the Berry phase, and $\delta$ is a phase shift depending on the dimensionality of the system [19]. The value of $\gamma$ $(= \frac{1}{2} - \frac{\phi_B}{2\pi} - \delta)$ provides valuable information about the topology of energy bands. For example, in 3D DSMs with linear energy dispersion, $|\gamma|$ is expected to lie in the range between 0 and 1/8, indicating the presence of a nontrivial $\pi$ Berry phase $\phi_B$ [20, 21]. According to the Onsager relation, the Landau level index $n$ should be a linear function of $1/B$, the slope and $n$-intercept of which yield the values of $\frac{A_F \hbar}{2\pi e}$ and $\gamma$, respectively. In Fig. 2 (c), the Landau fan diagram $n\left(\frac{1}{B}\right)$ at 1.6 K is plotted, with the peaks (or valleys) of the SdH oscillations indexed by integers (or half-integers). Indeed, the fan diagram can be well fitted linearly. The best linear fit of the fan diagram reveals a non-trivial Berry phase with $\gamma$=0.09. From the slope of the fitting curve, $A_F$=5.13e-4 Å$^{-2}$ is also obtained, similar to previous studies of $ZrTe_5$ [11].

According to the Lifshitz-Kosevich (LK) formula [19, 21], the SdH oscillation amplitude should decrease with increasing temperatures, which is described by the thermal damping factor $R_T = [(aTm^*/B)/sinh(aTm^*/B)]$, where $a = 2\pi^2 k_B/(e\hbar)$, $m^*$ is the effective mass, and $k_B$ is the Boltzmann constant. Therefore, from the analysis of the temperature dependence of the oscillation amplitude, one can derive the effective mass of the carriers. Fig. 2 (d) shows the fits of the normalized oscillation amplitude ($\Delta R_{xx}/\Delta R_{xx}(1.6\mathrm{K})$) of the $5^{th}$ landau level to the damping factor $R_T$. The fitting yields $m^* = 0.033m_0$, where $m_0$ is the free electron mass.

Therefore, the quantum oscillation results shown in Fig. 2 collectively reveal the presence of a band with a nontrivial Berry phase and a small effective mass in $ZrTe_5$. As discussed in previous transport studies, it is the conduction band at the $\Gamma$ point giving rise to the observed SdH oscillations [11, 13]. It is also noted that the existence of the nontrivial Berry phase is often regarded as a transport signature of the DSM phase in $ZrTe_5$ [11, 13]. But recent ARPES and infrared spectroscopy studies of $ZrTe_5$ suggest that $ZrTe_5$ is more like a weak or strong topological insulator with a small gap at the $\Gamma$ point at low temperatures, rather than the DSM

[16-18].

Fig. 2 (b) also shows a clear splitting of the 2nd LL at $T$=1.6 K, which was previously ascribed to the splitting of the Dirac node to two energetically-separated Weyl nodes in high magnetic fields [11,13]. As the temperature increases, the splitting-induced double-peak structure gradually evolves into a single peak at $T$=14.6 K, as seen in Fig. 2 (b). This reflects the smearing out of the splitting by the increased thermal energy at high temperatures. According to previous studies [23, 24], there are both Zeeman and orbital contributions to the exchange splitting induced by a magnetic field. The former mainly depends on the magnitude of the field and is isotropic with regard to the field direction, but the latter is highly anisotropic.

In order to discriminate the Zeeman and orbital contribution, we further studied the LL splitting in tilted magnetic fields at 1.6 K, as shown in Fig. 3 (a). $\theta$ is the angle of the field with respect to the $b$-axis in the $ab$ plane (see the inset). At small angles, the splitting of the 2nd LL is clearly resolved. But as $\theta$ increases, the two split peaks gradually emerge into a single one. In Fig. 3 (b), we plot the angle dependence of the spacing between the two peaks, *i.e.*, $\Delta_2 = 1/[(B_{2,+} - B_{2,-})cos\theta]$, where $B_{2,\pm}$ is the field position for the two split peaks of the 2nd LL, respectively. It can be seen that $\Delta_2$ almost keeps unchanged with increasing $\theta$, but beyond about 70°, a sharp drop of it to zero is observed. Such a strong angle dependence of $\Delta_2$ indicates the dominant anisotropic orbital splitting, rather than the isotropic Zeeman splitting. Similar orbital-dominant LL splitting has been also observed in some Dirac semimetals, such as $Cd_3As_2$ [24, 25]. Furthermore, it also demonstrates that the LL splitting in $ZrTe_5$ should be 3D-like, although $ZrTe_5$ has a layered structure. But in a previous study of $ZrTe_5$ [13], due to the limited range of $\theta$ ($\theta \leq 60^o$), the LL splitting is thought to be 2D-like, *i.e.*, the splitting only depends on the normal component of the field. Our study of the LL splitting in the extended range of $\theta$ clearly reveals the 3D character of it, as shown in Fig. 3 (b). We also note that the 3D nature of the band structure of $ZrTe_5$ was also confirmed in a recent laser ARPES study [18].

We also studied the MR of $ZrTe_5$ at higher temperatures. As shown in Fig. 4 (a), $ZrTe_5$ exhibits typical classical MR behaviors around 100 K, *i.e.*, the MR firstly increases with increasing fields and then tends to saturate in high magnetic fields. But as the temperature increases above 140 K, a non-saturating MR up to 14 T is observed. In order to demonstrate the change of MR with $T$, the MR ratio, defined by $R(B = 14\ \mathrm{T})/R(B = 0\ \mathrm{T})$, is plotted as a function of $T$ in Fig. 4 (b). Although the MR ratio is only about 230% at 100 K, it shows a rapid increase above 140 K. After a maximum value of about 820% is obtained at 170 ~180 K, the ratio decreases with increasing temperatures. Therefore, a large enhancement of MR is clearly observed around 170 K. Such an enhancement should be associated with the non-saturating behavior of MR shown in Fig. 4 (a).

It has been shown that when a gapless semiconductor with linear energy dispersion enters the quantum limit, *i.e.*, all carriers occupy the lowest LL in high fields, a linear and non-saturating MR is expected [26]. But this is apparently not applicable to our results, since the non-saturating MR in Fig. 4 (a) is observed at quite high temperatures (>140 K). Besides this quantum mechanism, inhomogeneity or mobility disorder can also give rise to a non-saturating MR in high fields [27]. But the observation of clear SdH oscillations in relatively low fields as shown in Fig. 2 indicates the high quality of our $ZrTe_5$ single crystals. After ruling out the above two mechanisms, we intend to understand what we observed in terms of the classical two-carrier transport model, which has been shown to result in extreme magnetoresistance (XMR) in topological semimetals with perfect compensation between hole and electron carriers, such as $WTe_2$ and $TaAs_2$ [28, 29].

To provide more evidence for this two-carrier transport model, we have also measured the Hall effect of our sample at high temperatures, as shown in Fig. 5 (a). At 100 K, the measured Hall resistivity curve $R_{yx}(B)$ reveals the dominance of $n$-type carriers in $ZrTe_5$. But with temperatures increasing to about 140 K, the slope of the $R_{yx}(B)$ curve in low fields changes from negative to positive, implying the participation of holes in the transport in addition to electrons. One can also see from Fig. 5 (a) that the hole contribution becomes more important with increasing temperatures and finally dominates the Hall effect up to 14 T at high temperatures. This Hall measurement clearly demonstrates the coexistence of electrons and holes in $ZrTe_5$, especially around 170 K, which coincides well with the temperature where the enhancement of MR is observed, as shown in Fig. 4 (b). Such a coincidence leads us to believe that it is the two-carrier transport mechanism that yields the non-saturating MR in Fig. 4 (a). We have also used the following two-band model (Eq. (1)) to fit the nonlinear $R_{yx}(B)$ curves

in Fig. 5 (b) [27-29]

$$\rho_{yx} = \frac{1}{e}\frac{(n_p\mu_p^2 - n_e\mu_e^2) + \mu_p^2\mu_e^2B^2(n_p - n_e)}{(n_p\mu_p + n_e\mu_e)^2 + \mu_p^2\mu_e^2B^2(n_p - n_e)^2}B \quad (1)$$

$$\rho_{xx} = \frac{1}{e}\frac{(n_p\mu_p + n_e\mu_e) + (n_e\mu_e\mu_p^2 - n_p\mu_p\mu_e^2)B^2}{(n_p\mu_p + n_e\mu_e)^2 + \mu_p^2\mu_e^2B^2(n_p - n_e)^2} \quad (2)$$

where $n_p$ $(n_e)$ & $\mu_p$ $(\mu_e)$ are the density and mobility of holes (electrons), respectively. The obtained fitting parameters at 170 K are $n_p$= 4.3e17 $cm^{-3}$, $\mu_p$=21900 $cm^2V^{-1}s^{-1}$, $n_e$=3.7e19 $cm^{-3}$, and $\mu_e$=24.5 $cm^2V^{-1}s^{-1}$, respectively. Compared with the high-density electrons, the low-density holes have much higher mobility. One can also see that, even in the two-carrier transport regime, our sample is away from the perfect electron-hole compensation. That is the reason why relatively small MR is obtained in our $ZrTe_5$ samples, compared with the XMR in $WTe_2$ and $TaAs_2$ [28, 29]. In Fig. 5 (b), the MR at 170 K can be also well described by Eq. (2) of the two-band model, further indicating the validity of the model in interpreting our results.

Note that the coexistence of holes and electrons in $ZrTe_5$ has been also revealed in previous ARPES or transport studies of $ZrTe_5$ [16-19]. Briefly speaking, there is a Fermi pocket at the $\Gamma$ point and another trivial electron pocket along the $YM$ edge in the first Brillouin zone of $ZrTe_5$ [16]. At low temperatures, only electrons from both two pockets contribute to the transport in $ZrTe_5$. But as the temperature increases above $T_p$ (=138 K), a Lifshitz transition occurs, leading to the change of the Fermi pocket at the $\Gamma$ point from electrons to holes [16-19]. This scenario explains the emergence of high-mobility holes above 140 K (~ $T_p$), as discussed in Fig. 5. Therefore, our MR and Hall results in Fig. 4 and 5 can be regarded as a direct consequence of the Lifshitz transition in $ZrTe_5$. But it is worth pointing out that besides the Lifshitz transition, the electron pocket along the $YM$ edge is also essential to the occurrence of the abnormal enhancement of MR shown in Fig. 4. Without this pocket, the two-carrier transport mechanism can not be realized in $ZrTe_5$.

Although the two-carrier transport under perfect electron-hole compensation can give rise to XMR in some materials, such as $WTe_2$ and $TaAs_2$, this is usually achieved only at very low temperatures [28, 29]. For example, the MR of $TaAs_2$ at 9 T can be as high as 1200000% at 2 K, but it will dramatically drop to only about 12 % at 300 K [29]. For practical application of XMR in spintronics, one must find a way to realize the perfect electron-hole compensation condition at room temperature. As enlightened by our results shown in Fig. 4, if we can use a gate to tune the Fermi level (or the electron and hole densities) of our sample, we may possibly shift the temperature region where the large enhancement of MR is observed towards higher temperatures, which would be important for potential application of $ZrTe_5$ in magnetic sensors.

In conclusion, the measured SdH oscillations clearly demonstrate the existence of a non-trivial band with tiny effective mass in $ZrTe_5$. Although $ZrTe_5$ has a layered structure, the Landau level splitting appearing in high fields reveals the 3D anisotropic nature of it, distinctive from previous transport studies. The Lifshitz transition induced two-carrier transport, as indicated by the nonlinear Hall effect, gives rise to an abnormal enhancement of MR at high temperatures, which might be enlightening in potential application of $ZrTe_5$ in spintronics.

**Acknowledgement**

This work was supported by the National Natural Science Foundation of China (No. 11574129 and 11874194), the National Key Research and Development Program of China (No. 2016YFA0301703), and Technology and Innovation Commission of Shenzhen Municipality (No. KQJSCX20170727090712763 and KQTD2016022619565991). L.Z. and R.A. contributed equally to this paper.

**Figure Captions**

**Figure 1**. Temperature-dependent resistance of $ZrTe_5$ single crystal, with a peak anomaly appearing at 138 K. Inset: Schematic diagram of Hall devices made from $ZrTe_5$.

**Figure 2.** (a) Magnetoresistance measured at different temperatures with the field in the $b$-axis. All curves have been offset vertically for clarity. (b) The extracted oscillatory component $\Delta R_{xx}$ at different temperatures, with the peak position indexed by integer number $n$. (c) The landau fan diagram. (d) The normalized temperature dependent oscillation amplitude of the $n$ = 5 peak, fitted by the damping factor of the Lifshitz-Kosevich formula.

**Figure 3**. (a) High-field SdH oscillations at different tilting angles ($\theta$). For clarity, the data has been multiplied by 2 for $\theta \geq 64.1^o$. (b) Landau level splitting $\Delta_2$ of the 2nd LL as a function of $\theta$. $\Delta_2 = 1/[(B_{2,+} - B_{2,-})cos\theta]$, where $B_{2,\pm}$ is the field position for the two split peaks of the 2nd LL, respectively.

**Figure 4.** (a) Magnetoresistance measured at high temperatures from 100 to 280 K. For clarity, the curves are offset vertically. (b) Temperature dependent magnetoresistance ratio obtained at $B$=14 T.

**Figure 5.** (a) Field-dependent Hall resistance measured at temperatures as indicated. (b) A two-band model fitting of the $R_{yx}(B)$ and $R_{xx}(B)$ curve obtained at 170 K.

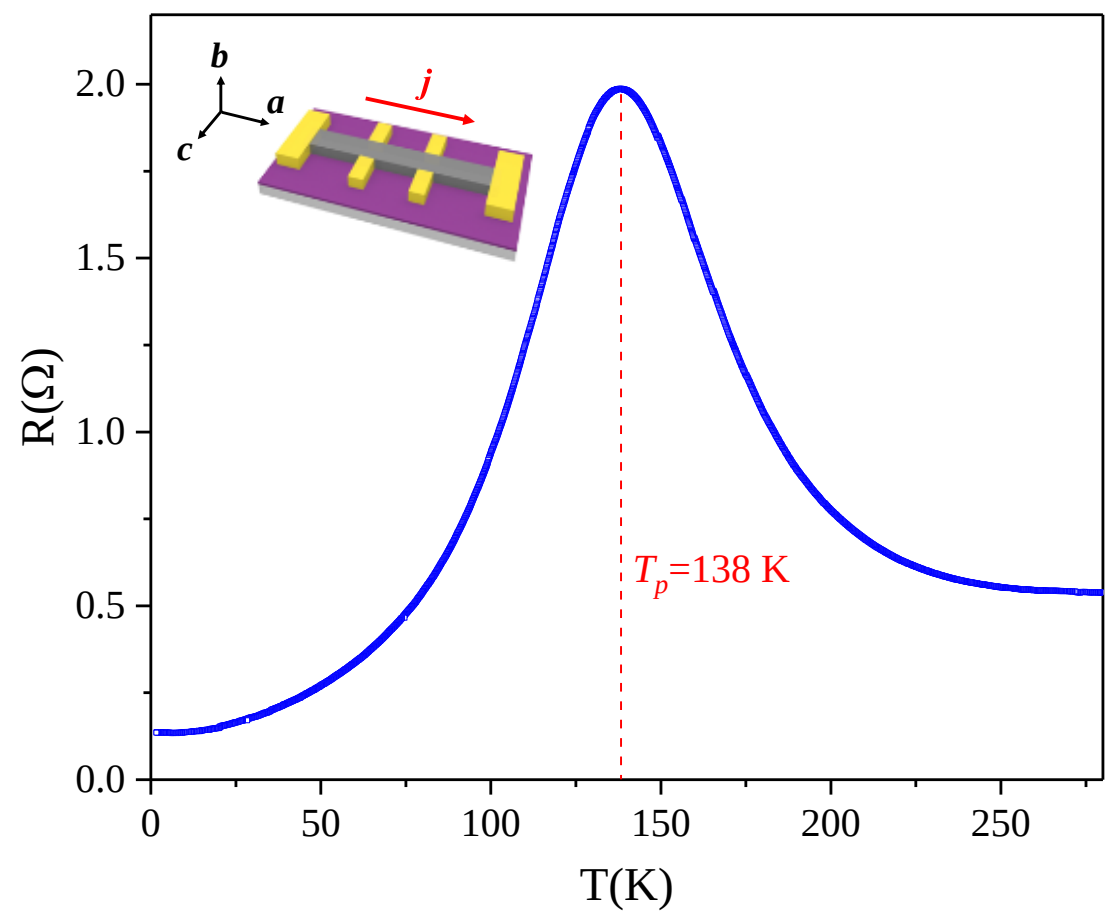

**Figure 1**

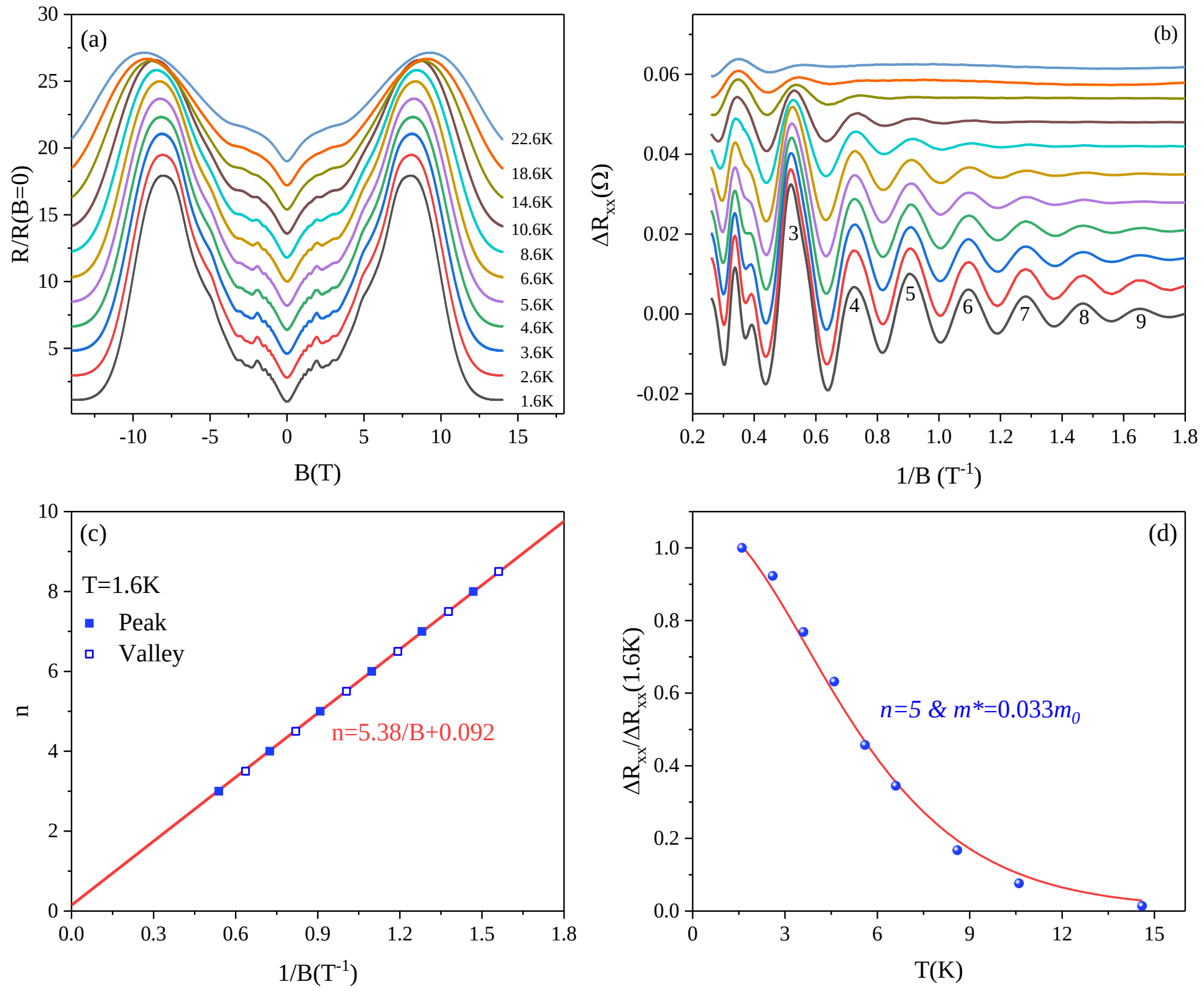

**Figure 2**

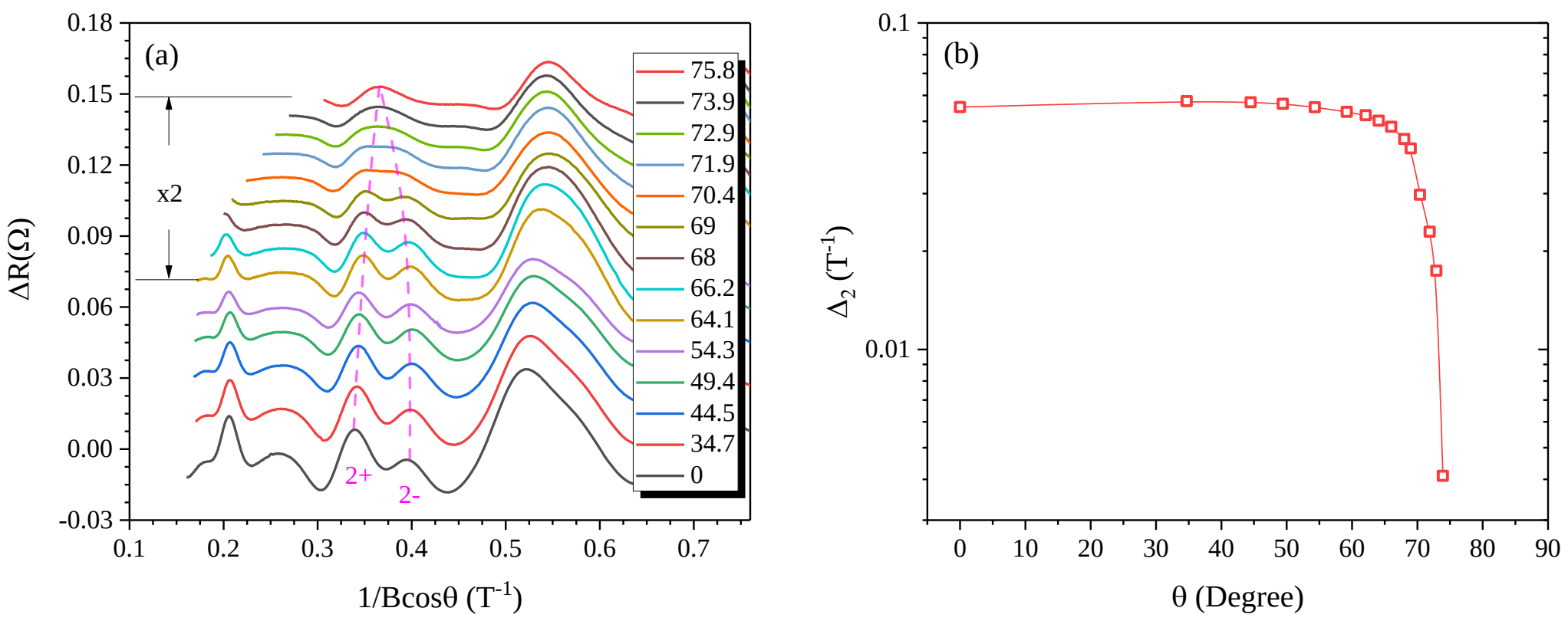

**Figure 3**

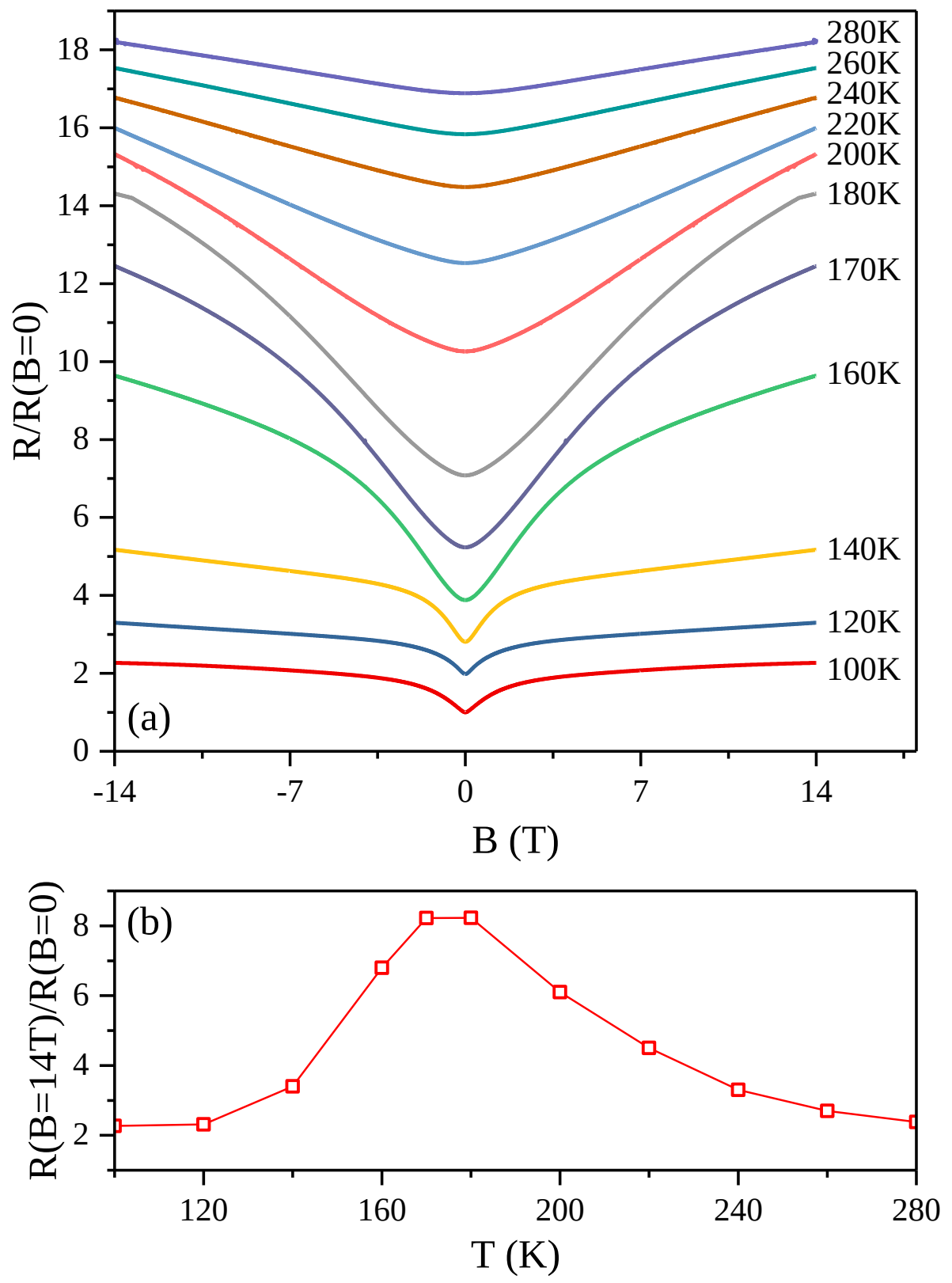

**Figure 4**

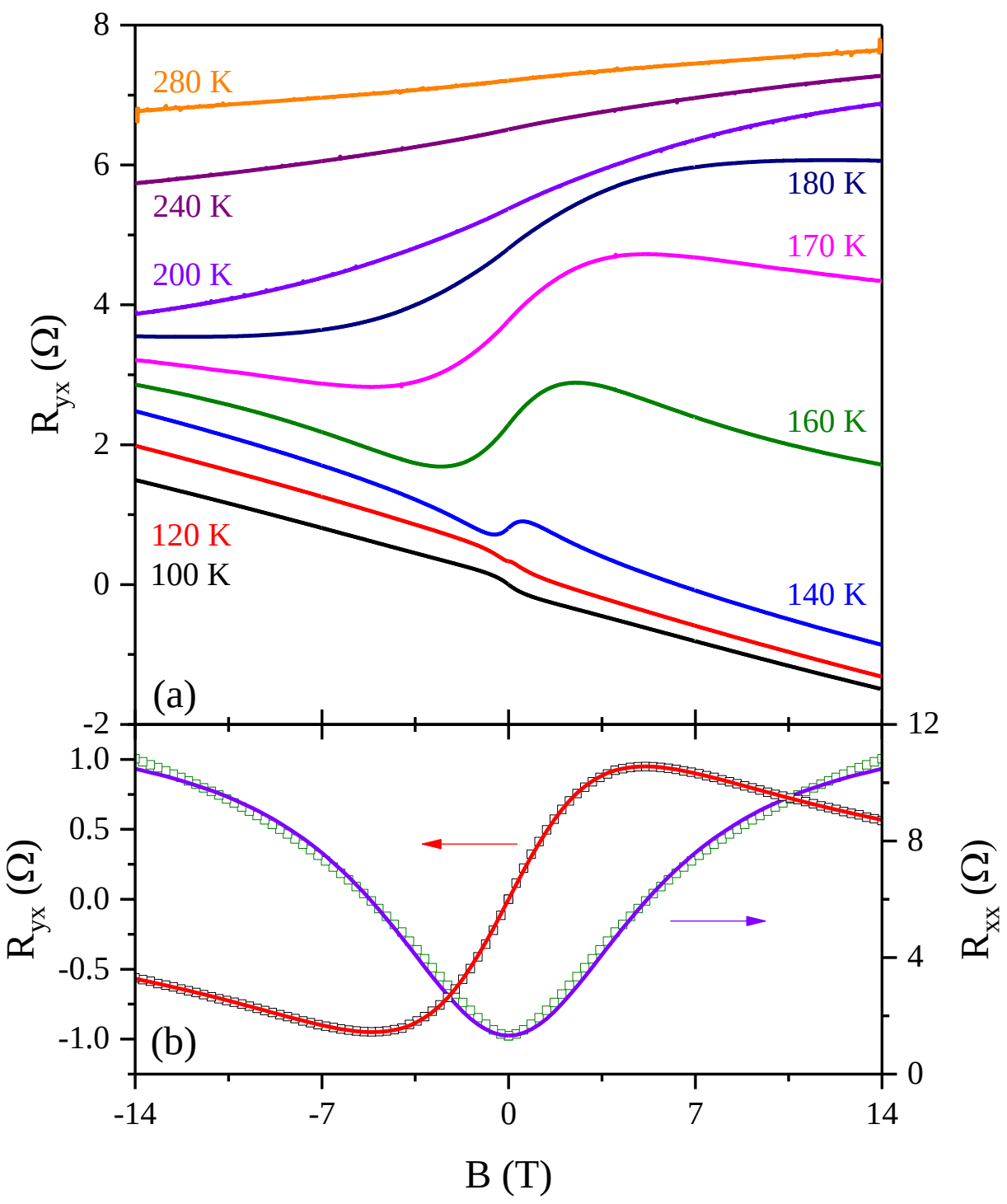

**Figure 5**